\documentclass{PoS}

\newcommand{\nc}[1]{\newcommand{#1}}

\nc{\its}[1]{\itshape #1 \upshape}
\nc{\mc}[3]{\multicolumn{#1}{#2}{#3}}

\nc{\bc}{\begin{center}}
\nc{\ec}{\end{center}}
\nc{\ig}[1]{\bc \includegraphics{#1} \ec}

\nc{\ben}{\begin{enumerate}}
\nc{\een}{\end{enumerate}}

\nc{\bo}[1]{\mbox{\boldmath \( #1 \! \! \)  \unboldmath}}

\nc{\be}{\begin{eqnarray}}
\nc{\ee}{\end{eqnarray}}
\nc{\bew}{\begin{eqnarray*}}
\nc{\eew}{\end{eqnarray*}}

\nc{\nnn}{\nonumber}
\nc{\f}[2]{\frac{#1}{#2}}
\nc{\td}[2]{\f{d #1}{d #2}}
\nc{\pd}[2]{\f{\partial #1}{\partial #2}}
\nc{\suli}{\sum\limits}
\nc{\proli}{\prod\limits}
\nc{\ili}{\int\limits}
\nc{\sr}[2]{\stackrel{#1}{#2}}
\nc{\dps}{\displaystyle}
\nc{\ket}[1]{\left| #1 \right>}
\nc{\bra}[1]{\left< #1 \right|}
\nc{\bracket}[2]{\left< #1 \right| \left. \! #2 \right>}
\nc{\norm}[1]{\left\| #1 \right\|}
\nc{\lndm}[1]{\pd{^{#1} \ln{\det{D}}}{\mu^{#1}}}
\nc{\pdmm}[1]{D^{-1} \pd{^{#1} D}{\mu^{#1}}}
\nc{\pdm}{D^{-1}\pd{D}{\mu}}
\nc{\trac}[1]{\mbox{Tr}\left(#1\right)}

\nc{\la}{\langle}
\nc{\ra}{\rangle}
\def\simge{\mathrel{%
       \rlap{\raise 0.511ex \hbox{$>$}}{\lower 0.511ex \hbox{$\sim$}}}}
\def\simle{\mathrel{
       \rlap{\raise 0.511ex \hbox{$<$}}{\lower 0.511ex \hbox{$\sim$}}}}
\nc{\tr}{{\rm tr \,}}
\nc{\Tr}{{\rm Tr \,}}
\nc{\Trc}{{\rm Tr_c \,}}
\nc{\trc}{{\rm tr_c \,}}
\nc{\bx}{{\bf x}}
\nc{\by}{{\bf y}}
\nc{\bz}{{\bf 0}}
\nc{\Te}{T_{\rm E}}
\nc{\Tpc}{T_{\rm pc}}
\nc{\mpmv}{m_{\rm PS}/m_{\rm V}}
\nc{\bU}{\overline{U} \! }
\nc{\opsi}{\overline{\psi}}
\nc{\oeta}{\overline{\eta}}
\nc{\Fl}{\hat{F}}
\nc{\muh}{\tfrac{\mu}{2}}
\nc{\nuh}{\tfrac{\nu}{2}}
\nc{\ah}{\tfrac{a}{2}}
\title{Transport coefficients of causal dissipative relativistic hydrodynamics in
quenched lattice simulations}

\ShortTitle{Transport coefficients of causal dissipative relativistic hydrodynamics
in quenched lattice simulations}

\author{\speaker{Yu Maezawa} \\
        Mathematical Physics Laboratory, RIKEN Nishina Center, Saitama 351-0198, Japan \\
        E-mail: \email{maezawa@ribf.riken.jp}}
\author{Hiroaki Abuki \\
        Department of Physics, Tokyo University of Science, Tokyo
        162-8601, Japan}

\author{Tetsuo Hatsuda \\
 Department of Physics, The University of Tokyo, Tokyo 113-0033, Japan, and \\
  Institute for the Physics and Mathematics of the Universe (IPMU), 
  The University of Tokyo,
  Chiba 277-8568, Japan}

\author{Tomoi Koide \\
 Frankfurt Institute for Advanced Studies, D-60438 Frankfurt am Main, Germany}

\abstract{
Transport coefficients of causal dissipative relativistic fluid dynamics (CDR) 
 are studied in quenched lattice simulations. 
CDR describes the behavior of relativistic non-Newtonian fluids 
 in which the relaxation time appears as a new transport coefficient 
 besides the shear and bulk viscosities. 
It was recently shown that these coefficients can be given 
 by the temporal-correlation functions of the energy-momentum tensors 
 as in the case of the Green-Kubo-Nakano formula. By using the new formula
  in CDR, we study the transport coefficients 
 with lattice simulations in pure SU(3) gauge theory. 
After defining the energy-momentum tensor on the lattice, 
 we extract a ratio of the shear viscosity to the relaxation time 
 which is given only in terms of the static correlation functions. 
The simulations are performed on $24^3 \times 4$--16 lattices 
 with $\beta_{_{\rm LAT}} = 6.0$, which corresponds to the temperature range of
  $0.5 \simle T/T_c \simle 1.8$, where $T_c$ is the critical temperature.
}

\FullConference{The XXVIII International Symposium on Lattice Filed Theory\\
                 June 14-19,2010\\
                 Villasimius, Sardinia Italy}

\begin{document}

%%%%%%%%%%%%%%%%%%%%%%%%%%%%%%%%%%%%%%%%%%%%%%%%%%%%%%%%%%%
\section{Introduction}

Relativistic fluid dynamics is an important model to understand
various collective phenomena in astrophysics and heavy-ion collisions, 
although its theoretical foundation has not yet been established \cite{koide_review}.
The relativistic Navier-Stokes theory is, 
for example, acausal and unstable 
and inadequate as
the theory of relativistic fluids. The reason is that
the irreversible currents (the shear stress tensor $\pi^{\mu\nu}$, 
the bulk viscous pressure $\Pi$ etc.) are linearly proportional
to the thermodynamic forces (the shear tensor $\sigma^{\mu \nu}$,
the expansion scalar $\theta$ etc.), with 
the proportionality constant named the shear viscosity 
coefficient $\eta$,
the bulk viscosity coefficient $\zeta$ etc. Thus, the forces have
an instantaneous influence on the currents, which obviously violates
causality and leads to instabilities.
These problems are solved by, for example, introducing retardation into the 
definitions of the irreversible currents, leading to
equations of motion for these currents which thus become
independent dynamical variables. 
The retardation effect is characterized by the relaxation time.
Theories of this type are called
causal dissipative relativistic fluid dynamics (CDR).
In CDR, the irreversible currents and the thermodynamic forces are no
longer in a simple linear relation,
and such fluids are called {\it non-Newtonian}.
As a consequence, the transport coefficients for CDR cannot be computed
with methods commonly used for {\it Newtonian} (Navier-Stokes) fluids, 
such as the Green-Kubo-Nakano (GKN) formula.

Recently, a new microscopic formula 
to calculate the transport coefficients of CDR  
from time-correlation functions was proposed 
\cite{knk,hkkr}. 
This formula reproduces the ordinary results when it is applied to 
the classical Navier-Stokes theory and the diffusion equation.
The consistency between this new formula and the results obtained from 
the Boltzmann equation was confirmed in Ref. \cite{dhkr,dkr}.
Since this formula is derived from quantum field theory, 
it will be applicable even to dense fluids, differently from the calculations based on 
the Boltzmann equation.

The purpose of the present study is to calculate the transport coefficients 
of CDR with lattice QCD simulations by using the new formula. 
The calculations of the transport coefficients, in general, contain 
 temporal-correlation functions which are very difficult 
  to estimate in lattice simulations \cite{lattice1,lattice2}.
Thus, as a first attempt, 
 we focus on a ratio between the shear viscosity 
 and the corresponding relaxation time, 
 $\eta/\tau_\pi$, which is given only by static correlation functions.
After defining the correlation functions between the energy-momentum tensor on the
 lattice,
 we calculate the ratio in quenched lattice simulations
 on $24^3 \times 4$--16 lattices 
 with $\beta_{_{\rm LAT}} = 6.0$, which corresponds to the temperature range 
$0.5 \simle T/T_c \simle 1.8$ where $T_c$ is the critical temperature.

This report
is organized as follows:
in section \ref{sec:CDR}, we introduce formulations of CDR and show
 that the ratio of transport coefficients can be expressed 
 in terms of
  static correlation functions between the energy-momentum tensors.
In section \ref{sec:EMlattice}, the energy-momentum tensor is 
 defined on the lattice by using a clover-shaped combination of gauge links.
Results of lattice simulations 
 are shown in section \ref{sec:sim}, and
 the summary is given in section \ref{sec:summary}.

%%%%%%%%%%%%%%%%%%%%%%%%%%%%%%%%%%%%%%%%%%%%%%%%%%%%%%%%%%%
\section{Causal dissipative relativistic fluid dynamics}
\label{sec:CDR}

We first choose gross variables which are necessary to 
extract the macroscopic motion of many-body systems.
If the chosen variables are not enough, the derived fluid dynamics will show 
unphysical behaviors, such as instability
and the divergent transport coefficients.

For ideal fluid, the energy-momentum tensor $T^{\mu\nu}$ 
is a function only of the energy density $\varepsilon$ and the fluid velocity $u^{\mu}$, which is normalized as $u^\mu u_\mu = 1$.
Then, by applying a Lorentz transformation and using the 
definition of the energy density and pressure $P$, we obtain 
$T^{\mu\nu} = (\varepsilon + P)u^\mu u^\nu - g^{\mu\nu} P$. 
Note that $P$ is calculated by the equation of state.
Since $T^{\mu\nu}$ is conserved, we have 
\begin{equation}
\partial_\mu T^{\mu\nu} = 0.
\end{equation}
This is the relativistic Euler equation.

For dissipative fluid, $T^{\mu\nu}$ cannot be expressed only by $\varepsilon$ and $u^{\mu}$.
We represent this additional component by another second rank tensor $\Pi^{\mu\nu}$. 
The most general $T^{\mu\nu}$ is, then, given by 
$ T^{\mu\nu} = (\varepsilon + P)u^\mu u^\nu - g^{\mu\nu} P +\Pi^{\mu\nu}$.
Conventionally, $\Pi^{\mu\nu}$ is expressed using the trace part 
$\Pi$ and traceless part 
$\pi^{\mu\nu}$ as $\Pi^{\mu\nu} = \pi^{\mu\nu} - (g^{\mu\nu} -u^\mu u^\nu) \Pi$. 
%\footnote{The heat conduction is neglected, because it finally disappears by using the 
%definition of the fluid velocity.}.
Finally $T^{\mu\nu}$ is expressed as
\begin{equation}
T^{\mu\nu} = (\varepsilon + P +\Pi)u^\mu u^\nu - g^{\mu\nu} (P+\Pi) + \pi^{\mu\nu},
\end{equation}
and $\Pi$ and $\pi^{\mu\nu}$ are the bulk viscous pressure and the shear stress tensor, 
respectively, satisfying the orthogonality condition $u_\mu \pi^{\mu\nu} = 0$.
In the traditional Landau-Lifshitz theory \cite{ll}, 
the viscous terms are induced instantaneously by 
the corresponding thermodynamic force:
\begin{eqnarray}
\Pi = -\zeta \theta , ~~~~~~~~~~~~
\pi^{\mu\nu} = 2 \eta \sigma^{\mu\nu}, 
\end{eqnarray} 
where $\zeta$ and $\eta$ are the bulk and shear viscosities, respectively.  
The thermodynamic forces $\theta$ and $\sigma^{\mu\nu}$ 
are defined by 
\begin{eqnarray}
\theta = \partial_\mu u^\mu , \ \ \
\sigma^{\mu\nu} 
= \frac{1}{2}\left( \partial^\mu u^\nu + \partial^{\nu} u^\mu 
- \frac{2}{3} (g^{\mu\nu} - u^\mu u^\nu) \theta \right) 
\equiv \Delta^{\mu\nu\lambda \delta} \partial_\lambda u_\delta .
\end{eqnarray}
When we use these definitions of the viscous terms, we obtain 
the relativistic Navier-Stokes equation. 
Because of the instantaneous production of the viscous terms, 
this equation contains 
sound propagations 
with infinite speed.

In order to solve this problem, 
the retardation effect is taken into account by introducing relaxation times 
$\tau_\pi$ for the shear stress tensor 
and $\tau_\Pi$ for the bulk viscous pressure, respectively.
Thus the viscous terms satisfying causality are 
given by 
\begin{eqnarray}
\tau_\Pi u^\mu \partial_\mu \Pi + \tau_\Pi \Pi \theta + \Pi 
= - \zeta \theta,  \label{eq_Pi}  \ \ \
\tau_\pi \Delta^{\mu\nu\lambda\delta} u^\alpha \partial_\alpha \pi_{\lambda \delta} 
+ \tau_\pi \pi^{\mu\nu} \theta + \pi^{\mu\nu} 
= 2\eta \sigma^{\mu\nu}, \label{eq_pi}
\end{eqnarray}
where $\tau_\Pi$ and $\tau_\pi$ are the relaxation times of $\Pi$ and $\pi^{\mu\nu}$, 
respectively.
Here the projection operator $\Delta^{\mu\nu\lambda\delta}$ 
is necessary to satisfy the orthogonality relation.
These are the equations of CDR.
One can easily check that the Navier-Stokes theory is reproduced in the 
vanishing relaxation time limit.
The second terms on the l.h.s. come from 
the (de)compression
of fluid cells which is important to implement stable numerical
calculations with ultra-relativistic initial conditions \cite{dkkm4}.

In fluid dynamics, 
transport coefficients are inputs which should be calculated 
from the underlying microscopic dynamics.
As was discussed in the introduction, we cannot apply the GKN formula to CDR.
The new formula is derived by using the projection operator method 
\cite{knk,dhkr}. 
The results are summarized as 
\begin{eqnarray}
&& \frac{\eta}{\beta (\varepsilon + P)} 
=\frac{\eta_{GKN}}{\beta^2 \int d^3 {\bf x} (\hat{T}^{0x}({\bf x}),\hat{T}^{0x}({\bf 0}))}, 
~~
\frac{\tau_{\pi}}{\beta}
=
\frac{\eta_{GKN}}{\beta^2 \int d^3 {\bf x} (\hat{T}^{yx}({\bf x}),\hat{T}^{yx}({\bf 0}))}, 
\label{taueta}\\
&& \frac{\zeta}{\beta (\varepsilon + P)} 
= 
\frac{\zeta_{GKN}}{\beta^2 \int d^3 {\bf x} 
(\hat{T}^{0x}({\bf x}),\hat{T}^{0x}({\bf 0}))}, 
~~
\frac{\tau_{\Pi}}{\beta}
= 
\frac{\zeta_{GKN}}{\beta^2 \int d^3 {\bf x} 
(\delta \hat{\Pi}({\bf x}),\delta \hat{\Pi}({\bf 0}))}, 
\label{tauzeta}
\end{eqnarray}
where $\hat{~}$ denotes operator, and  we define
$\hat{\Pi} \equiv \sum_{i=1}^3 \hat{T}^{ii}/3 - c^2_s \hat{T}^{00}$ and 
$\delta \hat{A} \equiv \hat{A} - {\rm Tr}[\rho_{eq}\hat{A}]$ with the equilibrium density matrix 
$\rho_{eq}$.
The inner product is defined by Kubo's canonical correlation,
\begin{equation}
(A,B) = \int^\beta_0 \frac{d\lambda}{\beta} {\rm Tr}[\rho_{eq} A(-i\lambda) B].
\end{equation}
Here $\eta_{GKN}$ and $\zeta_{GKN}$ are the shear and bulk viscosities of Newtonian fluids 
which are calculated using the GKN formula (or more precisely, using the Zubarev method). 
These quantities are given by the temporal (dynamical) correlation functions.

One can see that the new transport coefficients are still 
calculated from the GKN formula with the normalization factors, 
which are, on the other hand, given by the static correlation functions.
Thus, for example, 
the ratio of the shear viscosity and corresponding relaxation time is 
calculated only from the static correlation functions,
\begin{equation}
\frac{\eta}{\tau_\pi (\varepsilon + P)}
= 
\frac{ \int d^3 {\bf x} (\hat{T}^{yx}({\bf x}),\hat{T}^{yx}({\bf 0}))}{ \int d^3 {\bf x} (\hat{T}^{0x}({\bf x}),\hat{T}^{0x}({\bf 0}))}.
\label{eq:ept}
\end{equation}
In the leading order of the weakly interacting bose gas, 
 the above ratio becomes $\frac{\eta}{\tau_\pi (\varepsilon + P)}
=\frac{P}{\varepsilon + P}$ which 
  becomes zero ($\frac{1}{4}$) 
 at $T=0$ ($T \rightarrow \infty$) for massive bosons.
In the following, we focus on this ratio and calculate it 
in quenched lattice simulations.

%%%%%%%%%%%%%%%%%%%%%%%%%%%%%%%%%%%%%%%%%%%%%%%%%%%%%%%%%%%
\section{Energy-momentum tensor on the lattice}
\label{sec:EMlattice}

Let us consider the gluonic matter at finite $T$, and 
 define the energy-momentum tensor for the SU(3) gauge theory
  in Euclidean space-time
  as,
\be
T_{\mu\nu}(x) = 
  2 \tr \left[ F_{\mu\alpha}(x) F_{\nu\alpha}(x) \right]
 - \frac{1}{2} \delta_{\mu\nu} 
 \left( 1 + \frac{\beta(g)}{2g} \right)
 \tr \left[ F_{\rho \sigma}(x) F_{\rho \sigma}(x) \right]
, \label{eq:Tmn}
\ee
where the trace is taken over color indices,
 and $\beta(g)$ is a beta function on the lattice \cite{Boyd:1996bx}.
 In the standard approach,  
 the field strength tensor squared on the lattice (without the summation over
  Lorentz indices)  is defined from 
  the Hermitian part of the plaquette as
\be
a^4 {\rm tr} \, \left[ F_{\mu\nu}(x) F_{\mu\nu}(x)\right] + O(a^5)
=
\beta_{_{\rm LAT}} \left[ 1 - \frac{1}{3} {\rm Re} \, {\rm tr} \, U_{\mu\nu}(x)
\right]
,
\ee
where $\beta_{_{\rm LAT}} = 6/g^2$ is a lattice gauge coupling.
This is utilized to define e.g. the standard gauge action.
However this does not tell us anything about  the off-diagonal part of the 
 energy-momentum tensor, $T_{\mu\nu} (\mu \ne \nu)$.
 Therefore,  
 the following equality (valid only in the continuum theory with 
  full O(3) rotational symmetry) has been employed 
  to calculate the correlations of
   the energy-momentum tensor:
\be
\left\la T_{ij}(x)T_{ij}(y) \right\ra = 
\frac{1}{2} \left[ \left\la T_{ii}(x)T_{ii}(y) \right\ra
 - \left\la T_{ii}(x)T_{jj}(y) \right\ra \right]
, \ \ \ (i,j=1,2,3).
\ee
It was however realized recently that  
 this relation receives large errors
  due to lattice discretization \cite{Meyer:2009vj}.
 Moreover, it does not give us a clue to calculate 
 the correlation of $T_{i4}$ (the denominator of 
  the ratio in  Eq.~(\ref{eq:ept})) at finite $T$.

Alternative way to define the field strength 
would be to take the anti-Hermitian part of the plaquette,
\be
a^4 {\rm tr} \, \left[ F_{\mu \nu} (x) F_{\rho \sigma} (x) \right] + O(a^5)
 \equiv 
 - \frac{\beta_{_{\rm LAT}}}{24} {\rm tr} \,
 \left( \big[ Q_{\mu\nu}(x)      - Q_{\mu\nu}^\dag (x)      \big]
        \big[ Q_{\rho \sigma}(x) - Q_{\rho \sigma}^\dag (x) \big] \right)
,
\label{eq:FF-QQ}
\ee
which can be used both for the first and second terms of the right hand side of 
 Eq.(\ref{eq:Tmn}).
 Here we adopt a clover-shaped combination of the plaquette \cite{Luscher:1996sc}
\be
Q_{\mu\nu}(x) \equiv 
\frac{1}{4} \left[
U_{\mu\nu}(x) + U_{\nu -\mu}(x) + U_{-\mu-\nu}(x) + U_{-\nu\mu}(x)
\right]
,
\ee
to respect the space-time symmetry.
This definition naturally leads to $\langle T_{\mu\nu} \rangle =0$
 for $\mu \ne \nu$.  
 In our simulation, we use the energy-momentum tensor
 obtained from Eq.(\ref{eq:FF-QQ}).

In the Euclidean space-time, 
the Kubo's canonical correlation for  the energy-momentum tensors 
 appearing in Eq.~(\ref{eq:ept}) becomes a susceptibility
\be
G_{\mu\nu} (T) = \frac{T^2}{V} \left\la \left(
\int d^3 \bx \int_{0}^{1/T} d\tau  T_{\mu\nu} ( \bx, \tau) 
\right)^2 \right\ra_T
,
\ee
where we have used the translation invariance both in spatial and 
temporal directions, and 
  $\la \cdots \ra_T$ denotes the thermal average at temperature $T$.
 With  $T = 1/(aN_t)$,    $V = (aN_s)^3$ and 
 $\int d^3{\bf x} \int d\tau  \rightarrow  a^4 \sum_x$
  on the lattice, 
 we can rewrite the static susceptibility $G_{\mu\nu}$ 
 in the lattice unit with zero temperature subtraction as
\be
\label{eq:Gmn}
G_{\mu\nu} (T) = %\frac{1}{a^3}
  \left[ 
  \left\la \frac{1}{N_s^3N_t^2}    \left( \sum_x T_{\mu\nu} (x) \right)^2 \right\ra_T
- \left\la \frac{1}{N_s^3N_{t0}^2} \left( \sum_x T_{\mu\nu} (x) \right)^2 \right\ra_{T=0} \right]
,
\ee
where $N_{t0}$ means the temporal lattice size at $T=0$.

%%%%%%%%%%%%%%%%%%%%%%%%%%%%%%%%%%%%%%%%%%%%%%%%%%%%%%%%%%%
\section{Results of lattice simulations}
\label{sec:sim}

We perform 
quenched lattice simulations employing 
 a standard plaquette gauge action
 on a isotropic lattice of
  $24^3 \times N_t$ with $N_t = 4 - 16$.
The lattice coupling is taken to be $\beta_{_{\rm LAT}} = 6.0$, which
 corresponds to $a=0.093$ fm with
the Sommer scale $r_0=0.5$ fm \cite{Umeda:2008bd}.
The range of $N_t$ corresponds
to $T/T_c \sim 0.5$--1.8,
 where the critical temperature is located  
between $N_t=7$ and $N_t=8$.
The zero-temperature subtraction is performed
 with $N_t = 24$.
We generate pure gauge configurations 
 by the pseudo-heat-bath algorithm
  and measure 
  correlations
 using 1000--5000 configurations
   at every 1000 trajectories after thermalization.
Statistical errors are estimated by the jackknife analysis.

\begin{figure}[t]
  \begin{center}
    \begin{tabular}{cc}
    \includegraphics[width=73mm]{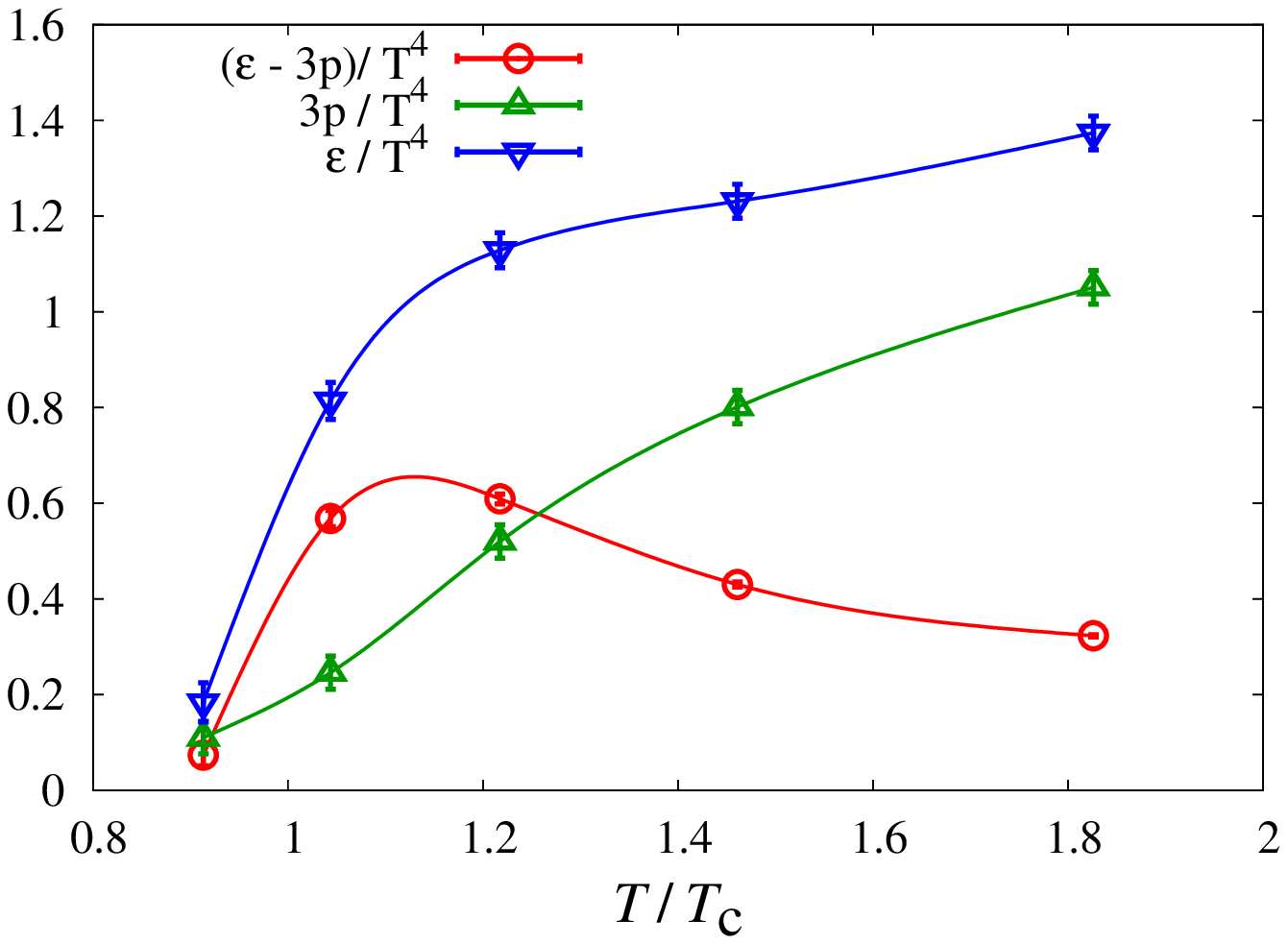} &
    \includegraphics[width=73mm]{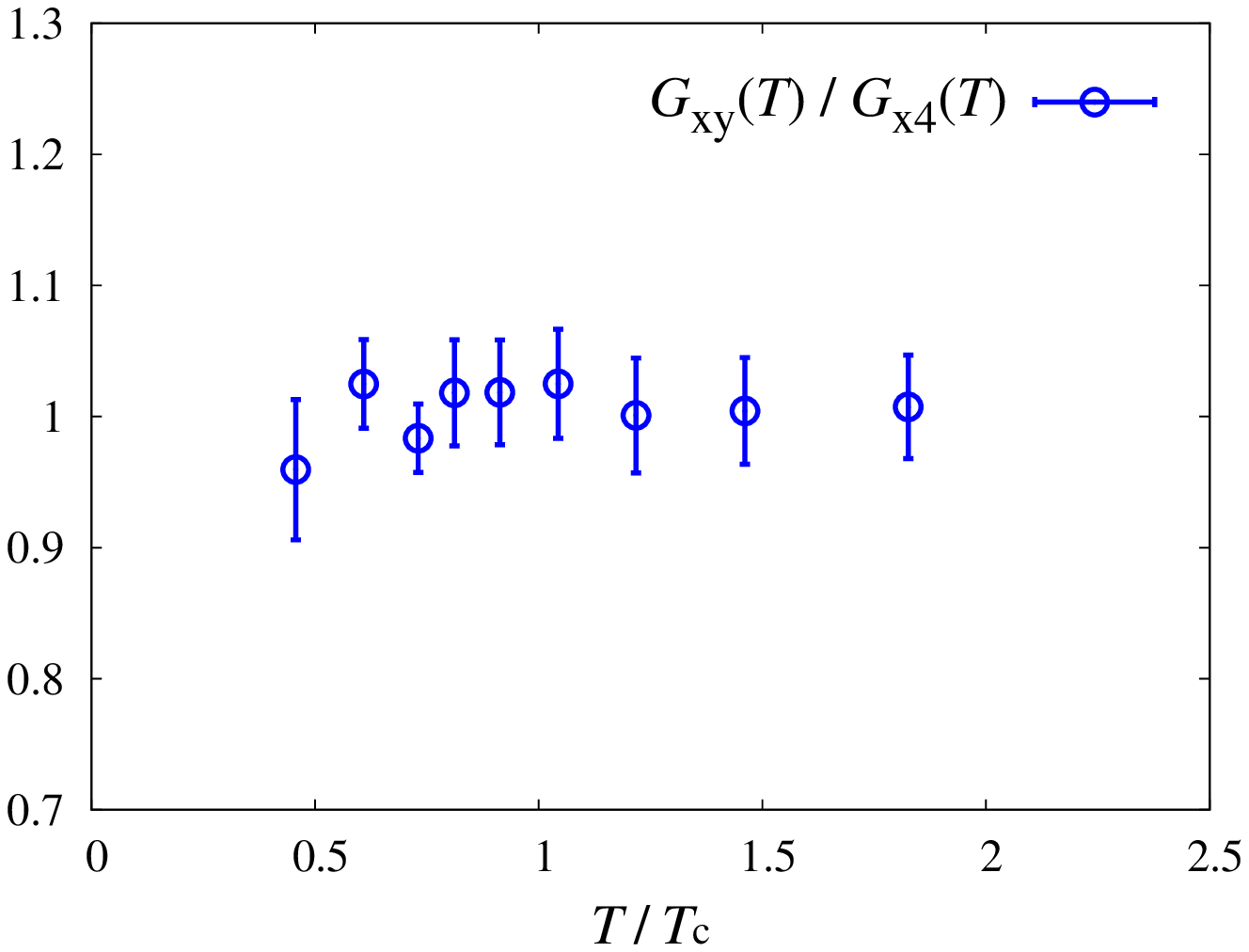}
    \end{tabular}
    \caption{Results of the trace anomaly, energy density and pressure (left)
    and the ratio of the susceptibilities $G_{xy}(T) / G_{x4}(T)$ (right)
     as a function of temperature.}
    \label{fig1}
  \end{center}
\end{figure}

In order to see the behavior of the energy-momentum tensor
 constructed from Eq.~(\ref{eq:FF-QQ}),
 let us first show results of the trace anomaly,
\be
\varepsilon - 3 P &=& 
 \left[ 
  \left\la \frac{1}{N_s^3 N_t   } \sum_x \sum_\mu T_{\mu\mu}(x) \right\ra_{T=0}
- \left\la \frac{1}{N_s^3 N_{t0}} \sum_x \sum_\mu T_{\mu\mu}(x) \right\ra_{T}
\right]
.
\ee
Figure \ref{fig1}(left) shows temperature dependence of the trace anomaly 
together with  the energy density and pressure calculated
  by the $T$-integral method \cite{Umeda:2008bd}.
 Typical  enhancement of $(\varepsilon - 3 P)/T^4$  around 
 $T_c$, and the rapid (slow) increase of the 
 energy density (pressure)  can be seen.
 The off-diagonal parts of the energy-momentum tensor
  are found to be zero within the statistical error, 
  $\la T_{\mu\nu} \ra_T \simeq 0, \ (\mu \ne \nu)$.

We define the averaged static susceptibilities $G_{xy}$ and $G_{x4}$
from Eq.(\ref{eq:Gmn}) as
\be
G_{xy}(T) \equiv \f{1}{3} \left( G_{12} + G_{13} + G_{23} \right)
, \ \
G_{x4}(T) \equiv \f{1}{3} \left( G_{14} + G_{24} + G_{34} \right)
.
\ee
From the simulation, we found that both
  $G_{xy}$ and $G_{x4}$ increase monotonically with temperature
 with similar values, so that
the ratio $G_{xy}/G_{x4}$ shown in Fig.~\ref{fig1} (right) corresponding to  
 $\eta/\tau_\pi(\epsilon+P)$ is almost 
unity over the range of temperatures we have explored,
    $0.5 \simle T/T_c \simle 1.8$.
 This behavior is in contrast to that  expected from the 
 weakly interacting bose gas mentioned at the end of
 sec.\ref{sec:CDR}, and is worth to be studied further.

%%%%%%%%%%%%%%%%%%%%%%%%%%%%%%%%%%%%%%%%%%%%%%%%%%%%%%%%%%%
\section{Summary}
\label{sec:summary}

We examined
transport coefficients of causal dissipative relativistic
 fluid dynamics (CDR) in quenched lattice simulations.
Based on the microscopic formulae proposed in 
Refs.~\cite{knk,hkkr},
  a ratio between
the shear viscosity 
and the corresponding relaxation time, 
 $\eta/(\tau_\pi(\varepsilon + P))$,  
was computed
 from the static correlation functions of the energy-momentum tensor.
 We calculated these static correlation functions in quenched lattice simulations 
 on $24^3 \times 4$--16 lattices with $\beta_{_{\rm LAT}} = 6.0$, 
 which correspond to the temperature range of $0.5 \simle T/T_c \simle 1.8$.
In this temperature region, the ratio stays constant and close to unity.

\section*{Acknowledgments}
The results of the numerical simulations were performed
 by using the RIKEN Integrated Cluster of Clusters (RICC).
T.H. was  supported in part by Grant-in-Aid  of the  Ministry of
Education, Science and Technology, Sports and Culture (No.
20105003).

\end{document}